# NUMERICAL CODE SELFAS-3 AND ELECTRODYNAMIC AGGREGATION OF MAGNETIZED NANODUST

A.B. Kukushkin, V.S. Neverov

RRC "Kurchatov Institute", Moscow 123182, Russia

**Abstract.**
The principles of the parallel numerical code SELFAS-3 are presented. The code modifies previous version of the code to enable parallel computations of electrodynamic aggregation in a many-body system of basic blocks which are taken as strongly magnetized thin rods (i.e., one-dimensional static magnetic dipoles), with electric conductivity and static electric charge, screened with its own static plasma sheath. The aggregation modelling includes the electric current dynamics in a complicated evolving network to describe the processes of external and internal electric short-circuiting. The code enables the continuous modelling of a transition between the following states: randomly situated ensemble of solitary basic blocks; electric current-carrying filamentary system; restructured filamentary network with a trend towards a fractal skeletal structuring. The latter trend is illustrated with generation of a bigger magnetic dipole in (i) homogeneous random ensemble between the biased electrodes in the presence of a plasma electric current filament and (ii) random ensemble around a straight linear nanodust filament with inhomogeneous distribution of the trapped magnetic flux along the filament.

## 1. Introduction

We analyze the capability of the model which has been suggested to be responsible for the unexpected longevity of straight filaments, and their networks, revealed in the Z-pinch gaseous electric discharges (see [1] and references therein). This hypothesis predicted the macroscopic fractal structures with basic topological building block of tubular form (with presumably carbon nanotube (**CNT**) at the nanometer length scales), which is successively self-repeated at various length scales (see also the surveys [2,3] and web pages [4,5]). The possibility of self-assembling of a fractal filamentary structure from a magnetized electroconductive nanodust was studied in [6-12]. We use the 3-D numerical model for a many body system of strongly magnetized thin rods (i.e. 1D static magnetic dipoles). Each block possesses the longitudinal electric conductivity and the electric charge, statically screened with its own plasma sheath. Numerical modeling of $\sim 10^2$-$10^3$ such dipoles has shown the possibility of electrodynamic self-assembling of a tubular skeletal structure from an ensemble of initially-linear artificially-composed filaments, linked to the biased electrodes. To substantiate such initial conditions, the possibility of self-assembling of quasi-linear filaments (and closing of the electric circuit) was studied for an initially random ensemble of basic blocks in the magnetic field of a plasma filament with internal longitudinal magnetic field (i.e., nearly force-free internal magnetic configuration).

Further substantial development of the hypothesis [13-15] for a fractal macroscopic skeleton, which repeats the CNT structure at larger length scales, was suggested in [16] and continued with a series of papers by several groups, where mechanical and electrophysical properties of this hypothetical, virtually-assembled nanomaterial, named in [16] as "super carbon nanotube", have been studied theoretically with various numerical methods (see recent papers [17,18]).

In this paper we continue studying theoretically the ways to *fabricate* a wide class of fractal skeletal nanomaterial (not as ideal fractal as super CNT but still a fractal) via *electrodynamic aggregation* of the above mentioned basic blocks with a strong contribution of internal self-organization processes (*self-assembling*). To this end, the code SELFAS-3 is designed and tested to enable parallel computations which continuously involve various stages of the assembling process, which formerly were treated separately.

Here we give a review of the object-oriented parallel (MPI) code SELFAS-3 (Sec. 2) and present new results on electrodynamic aggregation of magnetized nanodust (Sec. 3).

## 2. The model of electrodynamic aggregation of magnetized nanodust

We assume the elementary block to possess the following electrodynamic properties:
- the 1D static magnetic dipole;
- static positive electric charge;
- static longitudinal electrical conductivity.

Such a dipole may be formally represented as a couple of magnetic monopoles located on the tips of a rigid-body dipole. However, in the case strongly inhomogeneous and/or curvilinear magnetic field the force is not equal merely to the sum of the products of magnetic charge and local magnetic field strength; approximation of 1D structure assumes large length-to-diameter ratio for the block. For a tubular block, a magnetic dipole assumes trapping of magnetic flux by the walls due to such a high electric conductivity for circular currents, which prevents magnetic flux dissipation for a time needed for assembling of structures and, e.g., possible chemical bonding of the tips.

We put electric charge in the center of the rod and assume static screening by the ambient electrons at some screening radius $R_D$; electric charging and self-screening come from electron thermal and/or field emission from the tips of a 1D nanoparticle.

The above characteristics enable us to describe the following interactions of elementary blocks (see Fig. 1):
- mutual magnetic attraction and repulsion of the dipoles (i.e. interaction of circular electric current in the wall of one elementary block with similar current in another elementary block),
- action of external magnetic field on the magnetic dipole (i.e. interaction of circular electric currents in the walls of the elementary block with the external electric current, producing the magnetic field),
- screened electric repulsion of elementary blocks,
- action of magnetic field, produced by the longitudinal electric current in all the magnetic dipoles, on the given dipole (i.e. interaction of circular electric current in the walls of the elementary block with the longitudinal electric current in the walls of other blocks),
- interaction of longitudinal electric current in the walls of the blocks.

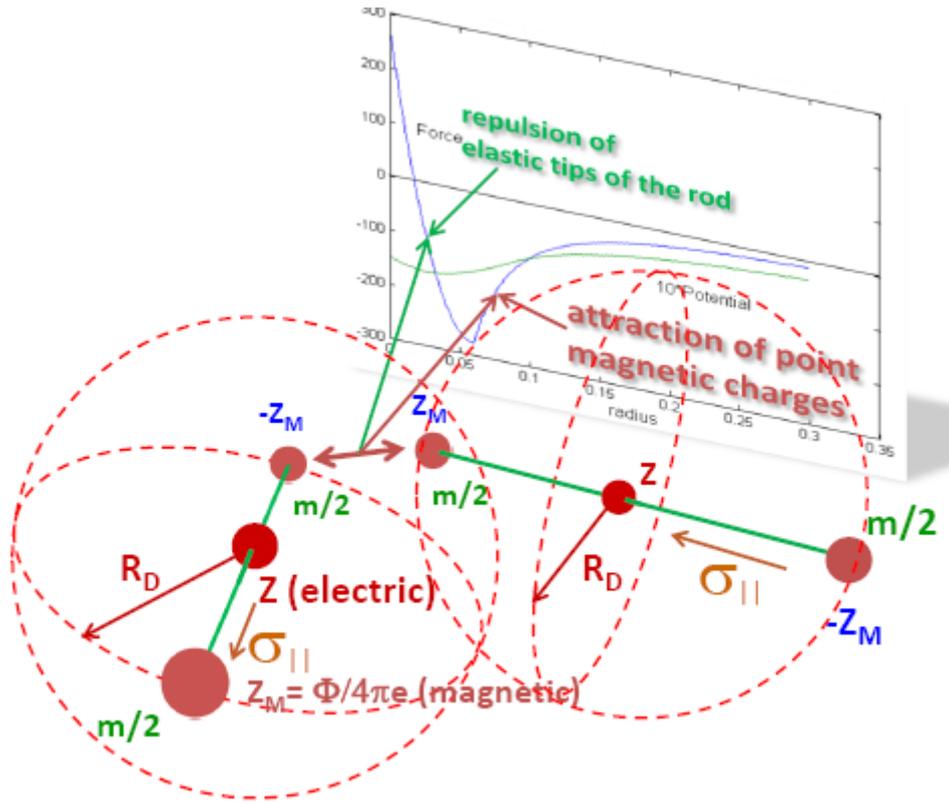

**Fig. 1.** Schematic drawing of the forces.

Once the formed filament short-circuits the baised electrodes or makes internal short-circuiting of an electric current-carrying nanodust filament, one has to recalculate (longitudinal) electric currents in a complicated filamentary network. We solve the respective circuit equations with neglect of the inductance and express the result for the current $J_i$ in the i-th block in the units of the current $J_0$ (see Eq.(11)) and in terms of parameter

$$\delta_{J0} = \frac{U L}{R_0 J_0 \Delta Z}, \qquad (1)$$

where $\Delta Z$ is the distance between electrodes, $U$ – voltage bias, $R_0$ – resistance of basic block, $L$ – elementary block's length (here we take the values $R_0$ and $L$ the same for all blocks to study, first of all, major trends in the aggregation process).

To simplify the description of dynamics of solid bodies we consider each dipole to be a couple of the point objects (coordinates $r_i$, masses $m_i$, i=1,2) which are linked together with a rigid-body massless bond and subjected to the action of the external forces applied to these objects, $F_1$ and $F_2$, and to the center of mass of the system (i.e. to the massless interconnecting bond), $F_{cm}$. The exact system of equations for such a system is described by the conventional equations for the motion of a solid body specified for the above particular case. The equations for the momentum and angular momentum, respectively, of the solid body are as follows:

$$\frac{\partial^2}{\partial t^2}\left(m_1 \vec{r}_1 + m_2 \vec{r}_2\right) = \vec{F}_1 + \vec{F}_2 + \vec{F}_{cm}. \qquad (2)$$

$$m_1 \left[ \vec{r}_1, \frac{\partial^2}{\partial t^2} \vec{r}_1 \right] + m_2 \left[ \vec{r}_2, \frac{\partial^2}{\partial t^2} \vec{r}_2 \right] = \left[ \vec{r}_1, \vec{F}_1 \right] + \left[ \vec{r}_2, \vec{F}_2 \right] + \left[ \vec{r}_{cm}, \vec{F}_{cm} \right] \quad (3)$$

where the square brackets denote the vector multiplication, and the radius vector of the center of mass of the system is equal to

$$\vec{r}_{cm} \equiv \left( m_1 \vec{r}_1 + m_2 \vec{r}_2 \right) / \left( m_1 + m_2 \right). \quad (4)$$

One may easily check that the solution to the system of Eqs. (2) and (3) may be found as a solution to the following system of equations:

$$m_1 \frac{\partial^2}{\partial t^2} \vec{r}_1 = -A \vec{r}_{12} + \vec{F}_1 + \frac{m_1}{m_1 + m_2} \vec{F}_{cm}. \quad (5)$$

$$m_2 \frac{\partial^2}{\partial t^2} \vec{r}_2 = A \vec{r}_{12} + \vec{F}_2 + \frac{m_2}{m_1 + m_2} \vec{F}_{cm}. \quad (6)$$

$$\vec{r}_{12} \equiv \left( \vec{r}_1 - \vec{r}_2 \right) \quad (7)$$

The first term in the right-hand side of Eqs. (5) and (6) describes the action of the rigid-body bond between the point objects 1 and 2. The value of A can be found from the condition that two rigidly bound point objects should have the same acceleration in the direction of their connectivity,

$$\left( \frac{\partial^2}{\partial t^2} \vec{r}_1 - \frac{\partial^2}{\partial t^2} \vec{r}_2, \vec{r}_1 - \vec{r}_2 \right) \equiv \left( \frac{\partial^2}{\partial t^2} \vec{r}_{12}, \vec{r}_{12} \right) = 0, \quad (8)$$

This gives

$$A = \frac{\mu_{12}}{r_{12}^2} \left( \frac{\vec{F}_1}{m_1} - \frac{\vec{F}_2}{m_2}, \vec{r}_{12} \right), \quad (9)$$

where $\mu_{12}$ is the reduced mass of the system of two point masses.

Major dimensionless variables of the problem outlined above are as follows. The space coordinates, time and velocity are taken in the units of dipole's length $L$, $t_0$ and $v_0$, respectively:

$$r_0 = L, \quad t_0 = \frac{\sqrt{mL^3}}{Z_{M0} e}, \quad v_0 = \frac{Z_{M0} e}{\sqrt{mL}}, \quad Z_{M0} = \frac{\Phi_0}{4\pi e}, \quad (10)$$

where $m = m_1 = m_2$, $Z_{M0}$ is the modulus of magnetic charge of the monopole taken in the units of electron charge $e$, $\Phi_0$ is characteristic value of magnetic flux trapped in the dipole, which, in our consideration, is basic parameter of dynamical problem. Electric charge Z is taken in the units of $Z_{M0}$.

All the forces are expressed in the units of $F_{M0}$ - magnetic attraction at the distance $L$. Longitudinal electric currents $J$ through the dipole and the external plasma current $J_{ext}$ are taken in the units of $J_0$, magnetic field is taken in the units of $B_0$:

$$B_0 = \frac{Z_{M0}e}{L^2}, \quad F_{M0} = \left(\frac{Z_{M0}e}{L}\right)^2, \quad J_0 = \frac{cZ_{M0}e}{L} \quad (11)$$

The electrodynamic forces are assumed to exceed largely the gravity force action on the dipoles.

To describe sticking of the dipoles we allow the magnetic monopoles to move freely in an isotropic potential well which is formed by
    (a) magnetic attraction of monopoles of the opposite sign and
    (b) their repulsion due to elasticity of the tips of the tubules/rods of finite diameter.
The form of the potential is shown in Figure 1. This potential provides smooth transition from the Coulomb potential for $r > r^*$ to a repulsion potential at smaller radii. Also, in the region $r < r^*$ we introduce the friction force:

$$\vec{F}_{brake} = k_{br}\vec{v}_{12}, \quad (12)$$

where $v_{12}$ is the relative velocity, and the coefficient $k_{br}$ is taken in the units of $k_{br0}$:

$$k_{br0} = \frac{\sqrt{m^3}}{Z_{M0}e\sqrt{L}}. \quad (13)$$

Also, the brake in an ambient medium is introduced in similar way to make possible the kinematic cooling of the formed structures via dissipation of the energy released in the close magnetic coupling ("magnetic recombination") of the dipoles.

The above strong simplification of the original picture of the motion of solid rods is acceptable if the spatial density of the rods is rather small and, respectively, the sticking and collisions of the rods are governed mostly by the interaction of strong magnetic monopoles on the tips of these rods.

The numerical method used to integrate equations (5)-(6) over time is the explicit 4-step Adams–Bashforth method. Under condition (8), to avoid continually growing error caused by the not infinitesimal value of time integration step, one has to correct new velocities obtained at each integration step to prevent the break-up or collapse of rigidly bound point objects. If $\vec{v}_{1A-B}^{n+1}$, $\vec{v}_{2A-B}^{n+1}$, $\vec{r}_{1A-B}^{n+1}$ and $\vec{r}_{2A-B}^{n+1}$ are new velocities and coordinates obtained from Adams–Bashforth integration, then the corrected velocities are as follows:

$$\vec{v}_1^{n+1} = \vec{v}_{1\,A-B}^{n+1} + v_{corr}\vec{r}_{12\,A-B}^{n+1}$$
$$\vec{v}_2^{n+1} = \vec{v}_{2\,A-B}^{n+1} - v_{corr}\vec{r}_{12\,A-B}^{n+1}$$
(14)

The value $v_{corr}$ is a solution of the equation:

$$(\vec{v}_1^{n+1} - \vec{v}_2^{n+1}, \vec{r}_{12\,A-B}^{n+1}) = 0 \tag{15}$$

Numerical model described above allows the increase of the time integration step up to 10 times in comparison with the previously used model in the same scenarios (cf. [6-12]). This model is implemented in the new object-oriented parallel (MPI) code SELFAS-3.

### 3. Numerical modeling of electrodynamic aggregation of magnetized nanodust

The code SELFAS-3 enables the continuous modelling of a transition between the following states: randomly situated ensemble of solitary basic blocks; electric current-carrying filamentary system; restructured filamentary network with a trend towards a fractal skeletal structuring. Here we present new results which illustrate these self-assembling processes. We make a stress at the process which is essential for the trend towards a fractal, namely generation of a bigger magnetic dipole. We illustrate it in the following two cases:

- **(i)** homogeneous random ensemble between the biased electrodes in the presence of a plasma electric current filament and an external magnetic field, both directed transverse to electrodes (Figs. 2-7),
- **(ii)** random ensemble around a straight linear nanodust filament, with inhomogeneous distribution of the trapped magnetic flux along the filament, in the presence of co-directed external magnetic field (here we take a model case of a single strong magnetic dipole, incorporated into a straight electric current filament; this corresponds to a local strong winding of the current which gives a thin solenoid on a straight current; Figs. 8-11).

The increase of block's number (cf. Fig. 2 and Fig. 6) substantially improves the identification of structuring in the calculation results.

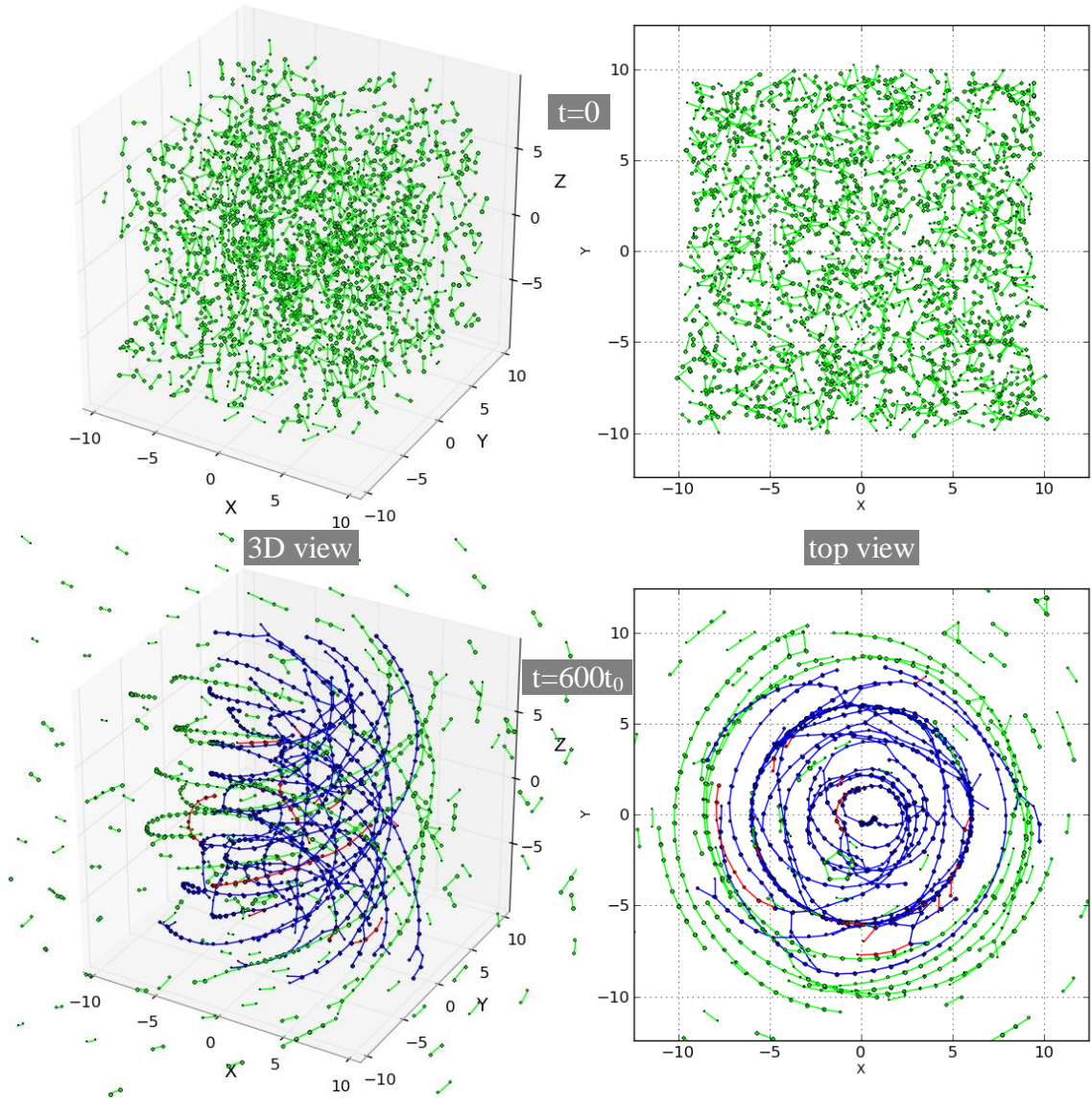

**Fig. 2.** Ensemble of 1200 basic blocks between electrodes at time $t = 0$ and $t = 600\ t_0$. Random initial spatial distribution of blocks. Short-circuited filaments are shown with blue color, while the dead ends within the network of short-circuited filaments are shown with red. Block's inverse aspect ratio $D/L = 0.06$, transition radius $r^* = D$ (tube's diameter), space dimensions of initial location region = [(-9.5,9.5);(-9.5,9.5);(-8.95,8.95)] (in units of block's length $L$), electric screening radius $R_D = L$, brake coefficients for tip-tip collision, $k_{br} = 100\ k_{br0}$, and for brake in an ambient medium, $M_{br} = 1.5\ k_{br0}$ (Eq. (13)), $\delta_{J0} = 2.23$ (Eq. (1)), external magnetic field $B_{ext} = [0,0,0.5]$ (in units of $B_0$ Eq. (11)), radius of (z-directed, with center at x=y=0) electric current filament $R_{plas} = 6\ L$, total longitudinal electric current through plasma filament $J_{zPlas} = 7.5\ J_0$. Electric charge $Z_i = Z_{Mi}$, while $Z_{Mi}$ is taken uniformly random in the interval $(2Z_{M0}/3, 2Z_{M0})$.

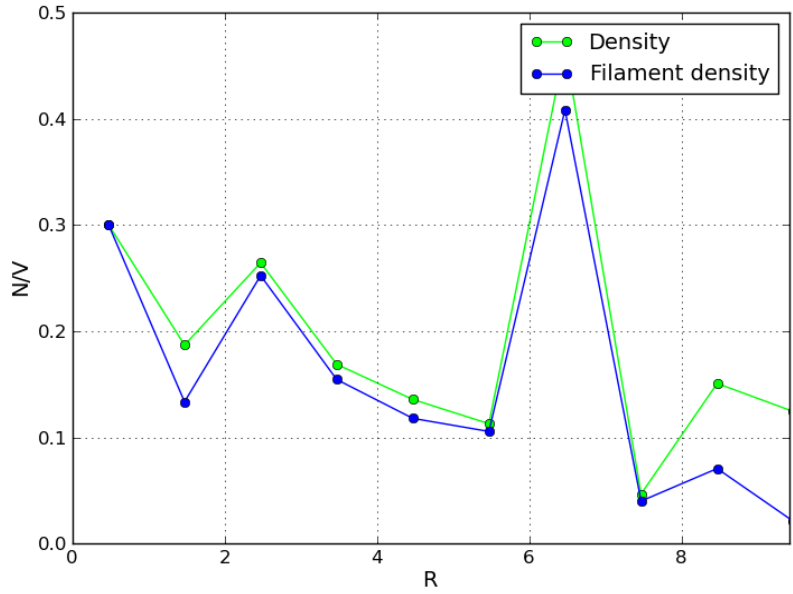

**Fig. 3.** Radial profile of density of basic blocks (averaged over longitudinal direction and azimuthal angle) at time $t = 600\ t_0$ for the system of Fig. 2. Blue – density of blocks in the short-circuited filaments. Green – total density.

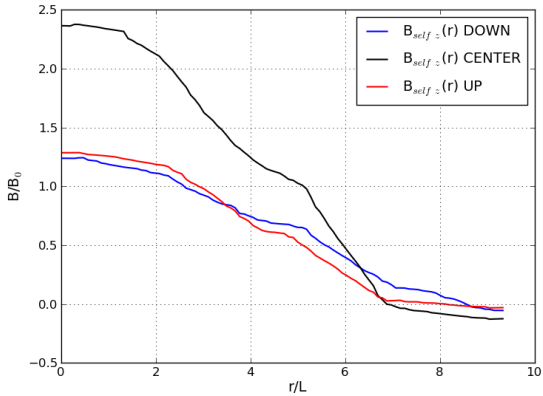

**Fig. 4.** Radial profile of Z-directed magnetic field produced by electric currents through the filaments composed of basic blocks, at time $t = 600\ t_0$ for the system of Fig. 2. Magnetic field is averaged over azimuthal angle.
Blue – in-plane $Z = -8.95\ L$,
Black – in-plane $Z = 0$,
Red – in-plane $Z = 8.95\ L$.

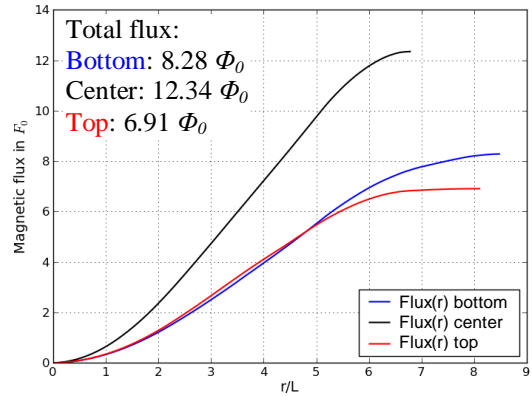

**Fig. 5.** Magnetic flux, calculated for the data from Fig. 4 and taken in units of $\Phi_0$, through circular cross-section in the XY plane.
Blue – in-plane $Z = -8.95\ L$,
Black – in-plane $Z = 0$,
Red – in-plane $Z = 8.95\ L$.

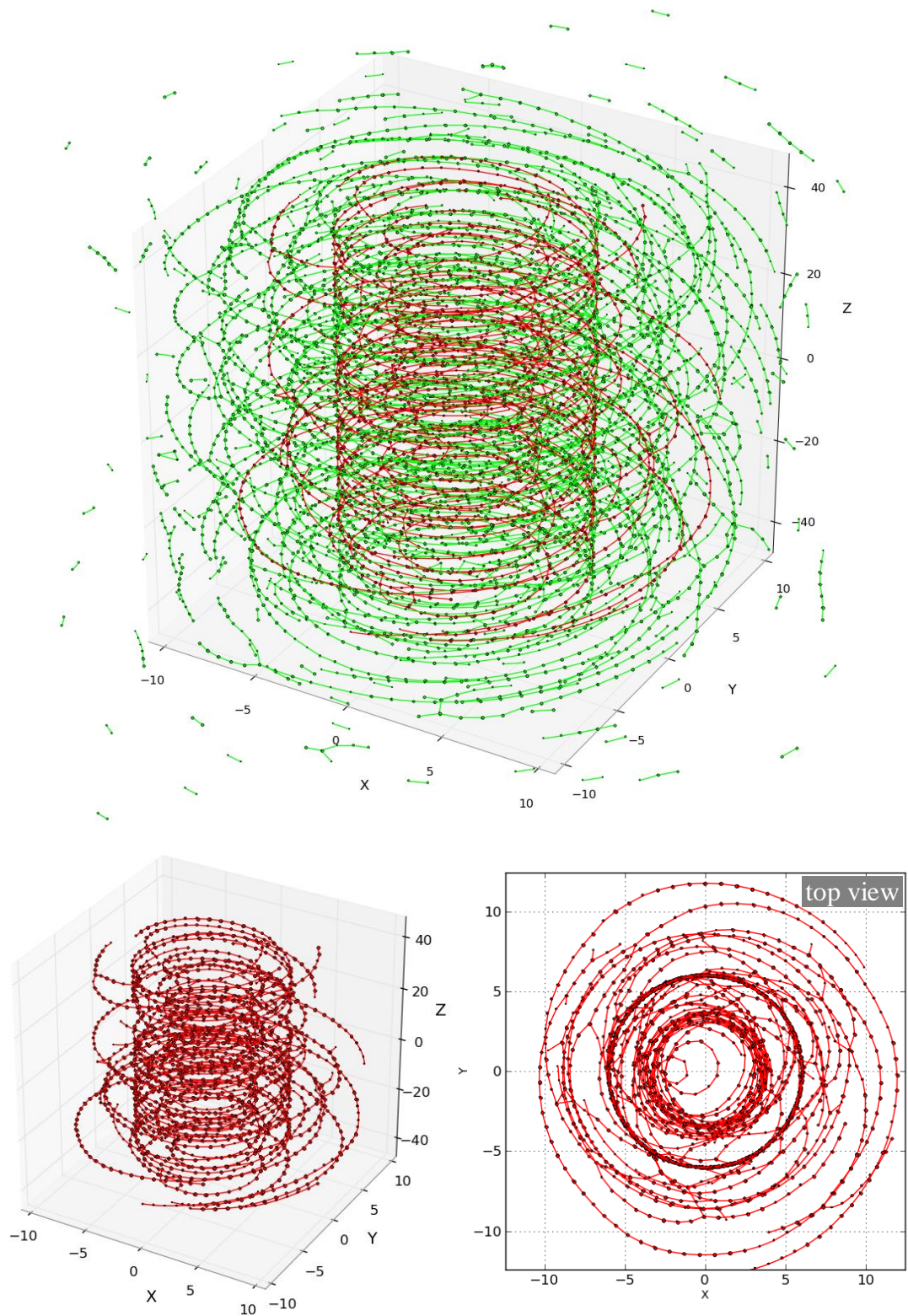

**Fig. 6.** Filamentary structuring in the ensemble of 6000 basic blocks at time $t = 100\, t_0$ for the conditions of Fig. 2 but with 5 times larger distance between the electrodes and 5 times greater value of parameter $\delta_{J0}$. Calculation was carried out on the 64 processor cores on the HPC cluster in the RRC «Kurchatov institute». Basic blocks, which form a single structure, are shown with red color.

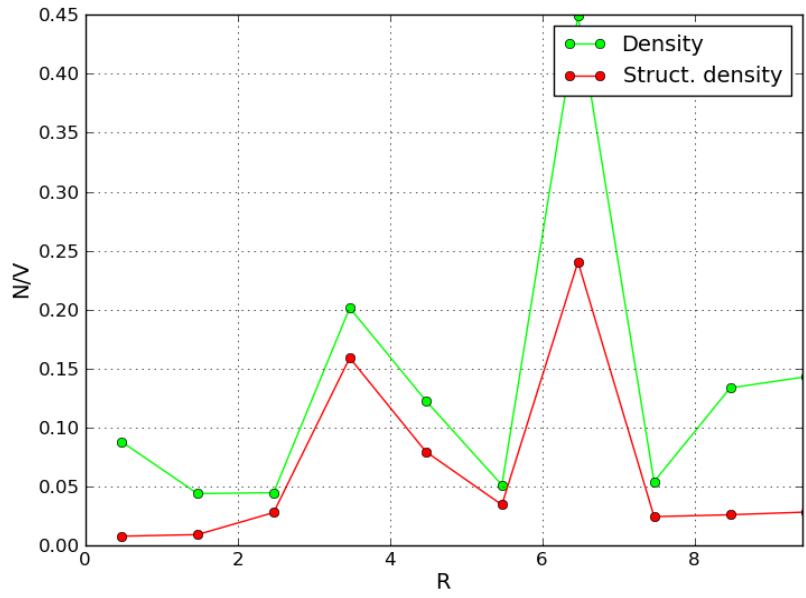

**Fig. 7.** Radial profile of density of basic blocks (averaged over longitudinal direction and azimuthal angle) at time $t = 100\,t_0$ for the conditions of Fig. 6. Red – density of the blocks in the single filamentary structure. Green – total density.

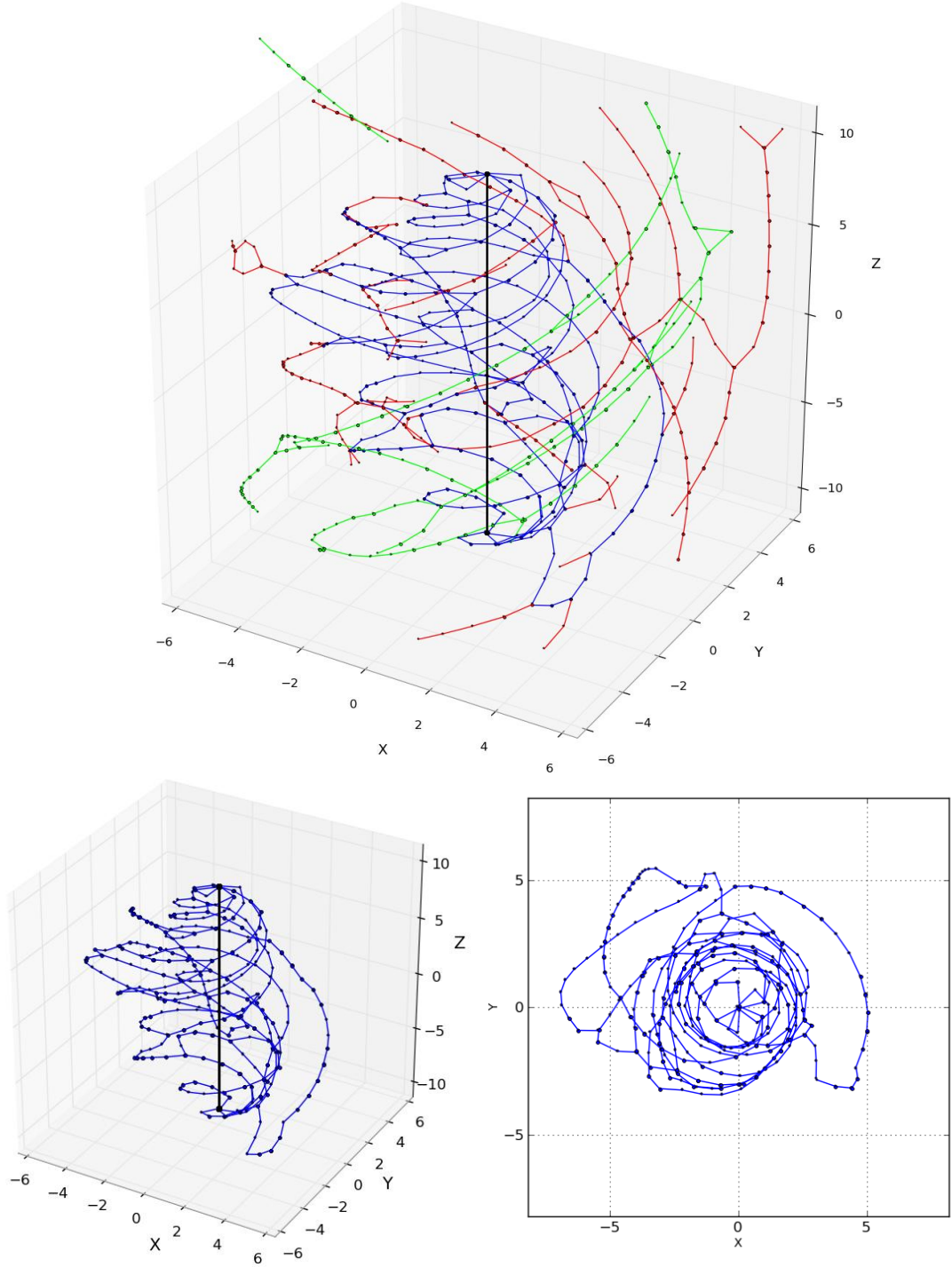

**Fig. 8.** Ensemble of 600 basic blocks around near external magnetic dipole (shown with black line), incorporated into a straight current filament (see Case (ii) ). Parameters of external dipole: $Z_{Mext}$ = 10 $Z_{M0}$, $L_{ext}$ = 20 $L$, electric charge density = 3.2 $Z_{M0}/L$, $R_{Dext}$ = 2.4 $L$, inverse aspect ratio $D_{ext}/L_{ext}$ = 0.009, longitudinal electric current in the filament $J_{ext}$ = 10 $J_0$, voltage bias corresponds to $\delta_{J0}$ = 2. Basic block's inverse aspect ratio D/L = 0.06, space dimensions of blocks' initial location region = [(-5.2,5.2);(-5.2,5.2);(-10,10)] (in units of block's length $L$), electric screening radius $R_D$ = 0.6 $L$, brake coefficients for tip-tip collision, $k_{br}$ = 100 $k_{br0}$, and for brake in a ambient medium, $M_{br}$ = 1.5 $k_{br0}$ (Eq. (13)), external magnetic field $B_{ext}$ = [0,0,1.5] (in units of $B_0$ Eq. (11)). Electric charge $Z_i = Z_{Mi}$, while $Z_{Mi} = 2Z_{M0}$ for 1/3 of blocks, and $Z_{Mi} = Z_{M0}$ for residual blocks. Short-circuited filaments are shown with blue color, while the dead ends within the network of short-circuited filaments are shown with red.

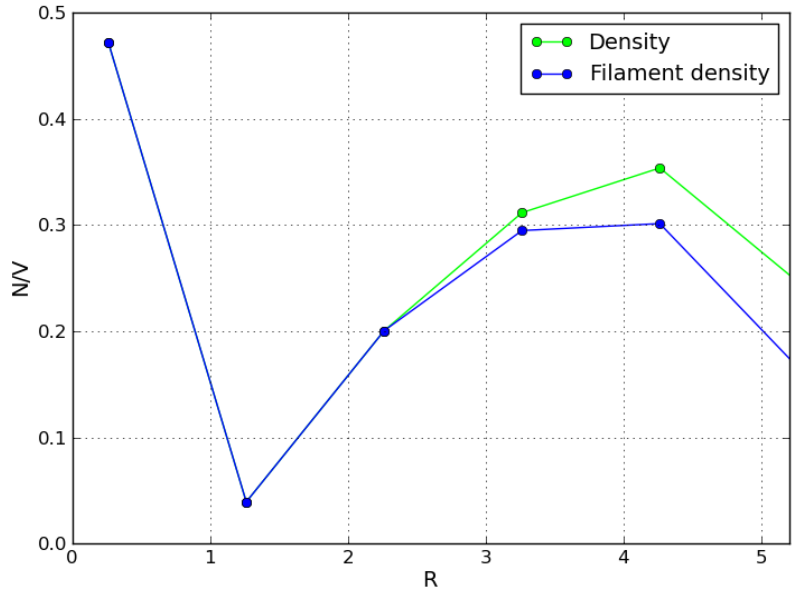

**Fig. 9.** Radial profile of density of basic blocks (averaged over longitudinal direction and azimuthal angle) at time $t = 400\ t_0$ for the conditions of Fig. 8. Blue – density of the blocks in electrical current structure. Green – total density.

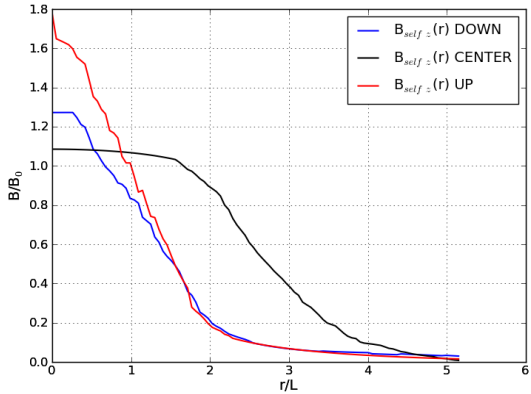

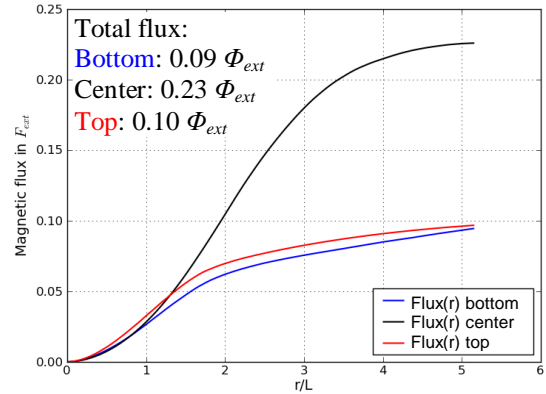

**Fig. 10.** Radial profile of Z-directed magnetic field produced by electric currents through the filament structure composed of basic blocks at time $t = 400\ t_0$ for the conditions of Fig. 8. Magnetic field is averaged over azimuthal angle.
Blue – in-plane $Z = -9.98\ L$,
Black – in-plane $Z = 0$,
Red – in-plane $Z = 9.98\ L$.

**Fig. 11.** Magnetic flux, calculated for the data from Fig. 10 and taken in units of $\Phi_{ext}$ (magnetic flux trapped by external large-scale magnetic dipole), through circular cross-section in the XY plane.
Blue – in-plane $Z = -9.98\ L$,
Black – in-plane $Z = 0$,
Red – in-plane $Z = 9.98\ L$.

## 3. Conclusions

1. The parallel numerical code SELFAS-3, which modifies previous version of the code to enable efficient parallel computations of electrodynamic aggregation in a many-body system of ($\sim 10^5$) strongly magnetized one-dimensional nanodust, makes it possible the continuous modelling of a transition between the following states: randomly situated ensemble of solitary basic blocks; electric current-carrying filamentary system; restructured filamentary network with a trend towards a fractal skeletal structuring. The increase of block's number substantially improves the identification of structuring in the calculation results.

2. The new results, obtained with SELFAS-3 for electrodynamic aggregation of nanodust, confirm the trend towards a fractal skeletal structuring. The latter is illustrated with generation of a bigger magnetic dipole in (i) homogeneous random ensemble between the biased electrodes in the presence of a plasma electric current filament and (ii) random ensemble around a linear nanodust filament with inhomogeneous distribution of the trapped magnetic flux along the filament.

**Acknowledgments**

The authors are grateful to I.B. Semenov, N.L. Marusov (RRC "Kurchatov Institute"), A.P. Afanasiev, M.A. Posypkin, A.S. Tarasov, V.V. Voloshinov (Institute for System Analysis RAS), for helpful discussions and a support of computational work.

This work is supported by the Russian Foundation for Basic Research (project RFBR 09-07-00469) and the European project EGEE-III (Enabling Grids for E-sciencE).